\documentclass[showpacs,twocolumn]{revtex4}

\usepackage[T1]{fontenc}
\usepackage[cp1250]{inputenc}
\usepackage{graphicx}% Include figure files
\usepackage{dcolumn}% Align table columns on decimal point
\usepackage{bm}% bold math
\usepackage{ulem}
\usepackage[centertags]{amsmath}

\begin{document}

\title{Magnetometry Based on Nonlinear Magneto-Optical Rotation with Amplitude-Modulated Light}

\author{S. Pustelny}
\author{A. Wojciechowski}
\author{M. Gring}
\author{M. Kotyrba}
\author{J. Zachorowski}
\author{W. Gawlik}
\affiliation{Centrum Bada\'n Magnetooptycznych, M.~Smoluchowski
Institute of Physics, Jagiellonian University, Reymonta 4, 30-059
Kraków, Poland}

\begin{abstract}
We report on an all-optical magnetometric technique based on
nonlinear magneto-optical rotation with amplitude-modulated light.
The method enables sensitive magnetic-field measurements in a broad
dynamic range. We demonstrate the sensitivity of
$4.3\times10^{-9}$~G/$\sqrt{\text{Hz}}$ at 10~mG and the magnetic
field tracking in a range of 40~mG. The fundamental limits of the
method sensitivity and factors determining current performance of
the magnetometer are discussed.
\end{abstract}

 \maketitle

\section{Introduction\label{sec:Introduction}}

Optical magnetometers explore optical signals which exhibit suitable
dependence on magnetic field. One category of optical magnetometers
applies high-resolution laser and/or radio-frequency (rf)
spectroscopy for generation of narrow resonances whose positions in
frequency vary with the magnetic field. Sensitivity of these devices
depends mainly on the width of a given resonance, whereas the
measurement range depends on the ability to determine the resonance
position. Second category of optical magnetometers exploits
magneto-optical phenomena, mainly the Faraday \cite{Faraday1846},
Macaluso-Corbino \cite{Macaluso1898}, and Hanle \cite{Hanle1924}
effects. These phenomena are characterized by changes of the
scattered-light intensity or polarization occurring around
zero-magnetic field. The resonances may have dispersive shapes which
allows to determine weak magnetic fields ($B\approx 0$) within the
range comparable to the resonance widths. However, a common
constraint of the devices from both categories is that the resonance
width affects not only the sensitivity but also the range of
measurable magnetic fields; the smaller the width, the better
sensitivity but narrower range.

Significant progress in the development of optical magnetometers
exploiting the magneto-optical rotation has been reached by the
advent of tunable lasers that allowed to take full advantage of
resonant enhancement of the Faraday rotation (see, for example,
Ref.~\cite{Gawlik1974CW}). Another important consequence of
application of lasers was the ability to explore the nonlinear
magneto-optical rotation (NMOR), particularly the nonlinear Faraday
effect associated with Zeeman coherences
\cite{Gawlik1974CW,ReviewNMOR}. The nonlinear Faraday effect
produced much narrower rotation resonances at $B=0$, which allowed
reaching sensitivity far greater than that achieved with
magnetometers employing the linear Faraday effect. However, the
increase of the sensitivity was associated with a considerable
reduction of the measurement range to very weak magnetic fields
only. A significant step in alleviating this drawback was the
application of synchronous pumping of atoms. This idea goes back to
the works of Bell and Bloom \cite{Bell1961} and Corney and Series
\cite{Corney1964} and the possibility of its application in context
of magnetometry leading to an extension of the dynamic range was
suggested before the ``laser era'' \cite{Tannoudji1970}. Yet, it was
only in Ref.~\cite{Budker2002FMNMOR} that this possibility was
experimentally verified. By the use of frequency modulation (FM) of
light \cite{Budker2002FMNMOR}, the rotation signal acquired multiple
extra resonances, the two most prominent occurring at
$B=\pm\hbar\Omega_\text{m}/2g\mu_\text{B}$, where $\Omega_\text{m}$
is the modulation frequency, $g$ the Land\'e factor, and
$\mu_\text{B}$ the Bohr magneton. The new, so-called, high-field
resonances are as narrow as the zero-field one but their positions
as a function of the magnetic field depend on the modulation
frequency, hence they can be used for ultra-precise measurements of
stronger fields. In a recent work \cite{Acosta2006High} the dynamic
range was extended up to the geophysical range with only a modest
loss of sensitivity caused by nonlinear Zeeman and ac Stark effects.

Although FM is a simple and convenient way of synchronous pumping,
necessary for observation of the new rotation resonances in non-zero
fields, it has its drawbacks, such as the ac Stark effect and
off-resonance pumping of atoms. On the other hand, the amplitude
modulation (AM) not only enables elimination of most of these
limitations but, additionally, offers a possibility to apply
specific modulation patterns. In particular, the pulsed excitation
and probing, either with a standard one-beam arrangement
\cite{Gawlik2006AMOR}, or with two separated beams (pump-probe
configuration) \cite{Higbie2006AllOptical}, appears to be a very
powerful magneto-optics tool for studies of the dynamic aspects of
magneto-optical phenomena. Low duty cycles allow observation not
only of the resonances at $B=\pm\hbar\Omega_\text{m}/2g\mu_\text{B}$
but also at multiplicities of these values. The AMOR (Amplitude
Modulated magneto-Optical Rotation) technique and its suitability
for creation of additional resonances extending the dynamic range of
magnetometric measurements have been first described in
Ref.~\cite{Gawlik2006AMOR}. In Ref.~\cite{Balabas2006Milllicells}
the AMOR technique was compared with the NMOR technique exploiting
FM of light.

In this paper an application of the AMOR technique to
high-sensitive, broad-dynamic range magnetometry is described. We
demonstrate application of the technique for magnetic field
measurements within the range of 0.1--40~mG with the sensitivity
$4.3\times 10^{-8}$~G/$\sqrt{\text{Hz}}$. Analysis of the
limitations of the present arrangement shows possibility of
extension of the measurement range to geophysical fields and
reaching the sensitivity close to the fundamental quantum limit.

The paper is organized as follows. In Sec.~\ref{sec:Setup} the
experimental apparatus is described. In the next section the NMOR
signals are presented and different factors limiting amplitudes and
widths of the signals, hence sensitivity of the method, are
discussed. Special attention is drawn to atomic collisions which are
recognized as one of the main mechanisms of relaxation in dense
vapors ($>10^{11}$~atoms/cm$^3$). Section~\ref{sec:Tracking} is
dedicated to a description of the method of magnetic-field tracking
and the special algorithm enabling broad dynamic-range measurements.
In Sec.~\ref{sec:Noise} performance of the magnetometer is
described, we comment on fundamental limits of the magnetic field
measurements and discuss the demonstrated sensitivity of the method.
Finally, conclusions are summarized in Sec.~\ref{sec:Conclusions}.

\section{Experimental setup\label{sec:Setup}}

\begin{figure}[h]
 \includegraphics[width=\columnwidth]{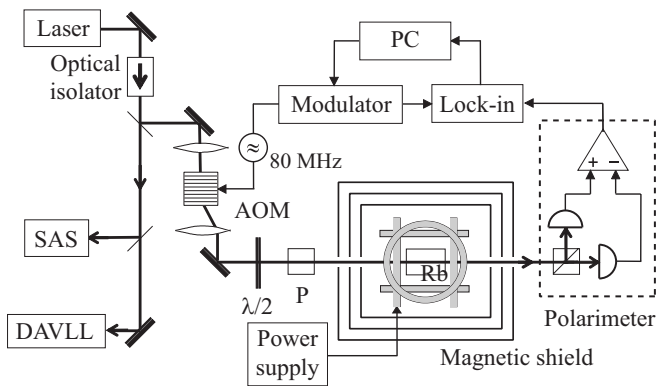}
 \caption{Experimental setup. SAS - saturated-absorption-spectroscopy frequency
 reference, DAVLL - dichroic-atomic-vapor laser lock, AOM - acousto-optical
 modulator, $\lambda/2$ - half-wave plate, P - polarizer.}
 \label{fig:Setup}
\end{figure}

The layout of the experimental apparatus is presented in
Fig.~\ref{fig:Setup}.
A sample of isotopically enriched $^{87}$Rb was contained in a
cylindrical vapor cell of 2-cm length and 1.8 cm in diameter. Inner
walls of the cell were coated with an antirelaxation (paraffin)
layer which prevented atoms from depolarizing collisions with the
walls and in this way lifetimes of ground-state Zeeman coherences
were prolonged by three orders of magnitude to about 20~ms. In our
cell at room temperature the lifetime is determined mostly by
collisions of the atoms with uncoated surfaces inside the cell stem.
The cell was placed inside a non-magnetic oven
temperature-stabilized between 15 and 60$^\circ$C. Three nested
$\mu$-metal layers provided magnetic field shielding with about
10$^4$ efficiency. Residual magnetic fields remaining inside the
inner-most layer were compensated by a set of three mutually
perpendicular magnetic-field coils also used for application of a
well-controlled field to the rubidium atoms.

An external-cavity diode laser was used as a light source. Its
frequency was tuned to the center of the $F=2\rightarrow F'=2$
hyperfine component of the rubidium D1 line (795~nm) and stabilized
with the dichroic-atomic laser lock
\cite{Corwin1998DAVLL,Wasik2002DFDL}. Light frequency reference was
provided by saturation spectroscopy. Before traversing the vapor
cell, light passed through an acousto-optical modulator (AOM)
optimized for the first-order diffraction. An $80$~MHz
radio-frequency (rf) signal driving AOM was amplitude-modulated with
frequencies $\Omega_\text{m}$ ranging from 100~Hz to 50~kHz, with
different modulation depths
$m=(I_\text{max}-I_\text{min})/I_\text{max}$, waveforms, and duty
cycles. In front of the cell a high-quality crystal polarizer was
placed to ensure pure linear polarization of the incident light beam
of 2 mm in-diameter. Light intensity was adjusted by a
half-waveplate situated in front of the polarizer. After traversing
the cell, polarization of the light beam was analyzed by a balanced
polarimeter consisting of a Glan prism with an axis set at
45$^\circ$ with respect to the incident-light polarization and two
photodiodes. A polarimeter differential signal was fed to a lock-in
amplifier and measured at the first harmonic of the modulation
frequency $\Omega_\text{m}$. A lock-in signal was stored on a
computer which also controlled the light-modulation frequency
$\Omega_\text{m}$ and the magnetic field inside the shield.

\section{Results \label{sec:Results}}

\begin{figure}[h]
 \includegraphics[width=\columnwidth]{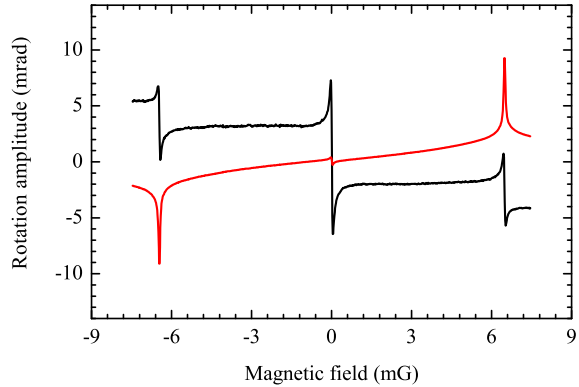}
\caption{(Color online) In-phase (black) and quadrature (red)
components of the NMOR signal recorded in the magnetic field domain.
The central feature seen at $B\approx 0$ is associated with the
``typical'' nonlinear Faraday effect while two satellites appear due
to synchronous pumping of atoms. The vertical offsets of the
high-field resonances in the in-phase component are related to
transit effect (see text). The signals were measured for
$I=8$~$\mu$W/mm$^2$ and square-wave modulation with 50\%
duty-cycle.} \label{fig:NMORMagneticFieldDomain}
\end{figure}

In Fig.~\ref{fig:NMORMagneticFieldDomain} the NMOR signal measured
versus the magnetic field with fixed modulation frequency
($\Omega_\text{m}\approx 1$~kHz) is presented.
As seen, the in-phase component of the signal consists of three
dispersive-like resonances. The central feature situated at $B=0$ is
associated with the ``typical'' nonlinear Faraday effect, i.e., a
similar signal would be observed with CW light \footnote{Note that
in an experimental scheme in which phase-sensitive detection of the
magneto-optical rotation is used, appearance of NMOR resonances is
only possible if light is modulated. For low magnetic fields
($\Omega_\text{L}\lesssim\gamma$) the modulation is not caused by
the Faraday rotation but need to be introduced externally. Since in
our experiment the same beam was used for pumping and probing, the
light was amplitude modulated and hence the lock-in detection of the
zero-field resonance was possible. On the contrary, in two-beam
experiment, in which separated beams are used form pumping and
probing atoms and probe-light intensity is unmodulated, hence the
zero-field resonance is not observed.}. Two satellite features are
the high-field resonances centered at
$B=\pm\hbar\Omega_\text{m}/2g\mu_\text{B}$. The widths of these
resonances, as well as the width of the zero-field resonance, are
determined by an effective time of interaction between light and
atoms. In paraffin-coated cells this time is related to the
so-called wall-induced Ramsey effect \cite{Kanorskii1995Wall};
polarized atoms leave the light beam, bounce off the walls many
times and return to the beam being still polarized. Thus the
effective time of interaction between light and atom is prolonged to
the time of round trip in which atomic polarization is preserved.

Two high-field resonances recorded in-phase are vertically offset in
opposite directions symmetrically relative to zero. The exact
mechanism of this offset is not clear yet and is a subject of
independent studies. Most likely, it is caused by atomic thermal
motions and atoms' transit across the light beam. In that case, the
atoms spend finite time within the light beam and do not reach
stationary conditions
\cite{Gawlik1986Nonstationary,Pfleghaar1993Time,Taichenachev2004Nonlinear,Alipeva2003Narrow}.
This leads to appearance of broad, not necessarily Lorentzian
contributions to the NMOR resonances with widths determined by the
transit time. In the described experiment this time was about
10~$\mu$s, hence its contribution to the NMOR signals was about
thousand times broader than the narrowest NMOR resonances.

The quadrature component of the signal consists of two high-field
absorptive-like features of opposite signs composed of narrow
resonances with broad pedestals. We associate the pedestals with the
transit effect, mentioned above. Another feature observed in
quadrature is a small, dispersive-like signal at $B=0$. Basing on
previous measurements \cite{Balabas2006Milllicells} and the observed
light-intensity dependence of that feature, we link it with the
alignment-to-orientation conversion \cite{Budker2000AOC}. This
effect arises when strong light converts atomic alignment into
longitudinal orientation by a combined action of the magnetic field
and electric field of light on the atomic polarization. For lower
light intensities the feature disappears which supports our
interpretation.

\begin{figure}[h]
  \includegraphics[width=\columnwidth]{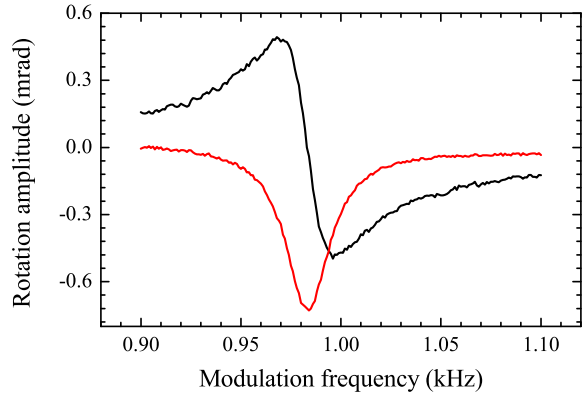}
  \caption{(Color online) In-phase (black) and quadrature (red) components of the NMOR
  signal as functions of the modulation frequency ($B\approx 0.7$~mG).
  The inset shows the vertical offset of the resonance, which is related to the transit effect
  (see text), for different magnetic fields. The signals were recorded for $I=5$~$\mu$W/mm$^2$
  and 50\% duty-cycle square-wave modulation.}
  \label{fig:ModulationFrequencySignal}
 \end{figure}

In Fig.~\ref{fig:ModulationFrequencySignal} the NMOR signal around
the high-field resonance is presented in the modulation-frequency
domain with fixed magnetic field.
As before, the in-phase and quadrature components are characterized
by the dispersive- and absorptive-like curves, respectively, that
are centered at 2$g\mu_\text{B}B/\hbar$ with the width determined by
the relaxation rate $\gamma$. Similarly to
Fig.~\ref{fig:NMORMagneticFieldDomain}, the in-phase component of
the NMOR signal is slightly offset. The offset depends on the Larmor
frequency, as shown in the inset, and it is characterized by a wing
of dispersive curve with the maximum at $\sim$16~kHz. This value is
consistent with the $\sim$10~$\mu$s time of atom flight through the
beam which confirms our interpretation of the offset.

Since the width of the NMOR signal is associated with the
ground-state relaxation rate, NMOR can be employed for
investigations of relaxation processes of the coherences
\cite{Pustelny2006Influence,Budker2005Antirelaxation}. Several
different physical mechanisms are recognized to be responsible for
ground-state relaxation in paraffin-coated cells. Some of them, such
as collisions with the walls or relaxation due to magnetic field
inhomogeneities, are independent of atomic concentration, while
others, e.g., collisions between the atoms, are density dependent.
The collisional relaxation is given by
\begin{equation}
\gamma_\text{c}=R(I) N\overline{v}\sigma,
 \label{eq:TotalRelaxation}
\end{equation}
where $N$ is the number density of atoms,
$\overline{v}=4\sqrt{k_\text{B}T/\pi m}$ is the the average relative
velocity, $k_\text{B}$ is the Boltzmann constant, $T$ is the
temperature, $m$ is the atomic mass, $\sigma$ is the collisional
relaxation cross-section, $R(I)$ is the so-called nuclear slow-down
factor \cite{Happer1972Review}, and $I$ is the nuclear spin.

In order to calculate the temperature dependence of the atomic
density, and hence the density dependence of the relaxation rate
$\gamma_\text{c}$, an analysis similar to that of
Ref.~\cite{Pustelny2005Density} is performed. We use the
phenomenological relation between pressure $p$ and temperature $T$
\cite{Nesmeyanov1963}
\begin{equation}
 \log_{10}p(T)=A-\frac{B}{T}+CT+D\log T,
\label{eq:PressureRelation}
\end{equation}
where $A$, $B$, $C$, and $D$ are the numerical factors, and the
ideal gas equation
\begin{equation}
 p(T)=Nk_\text{B}T.
 \label{eq:IdealGasEquation}
\end{equation}
Combining Eqs.~(\ref{eq:PressureRelation}) and
(\ref{eq:IdealGasEquation}) one obtains relation between $N$ and $T$
\begin{equation}
 N=\frac{10^{A-B/T+CT+D\log T}}{k_\text{B}T}.
 \label{eq:ConcentrationDependence}
\end{equation}

\begin{figure}[tb]
   \includegraphics[width=\columnwidth]{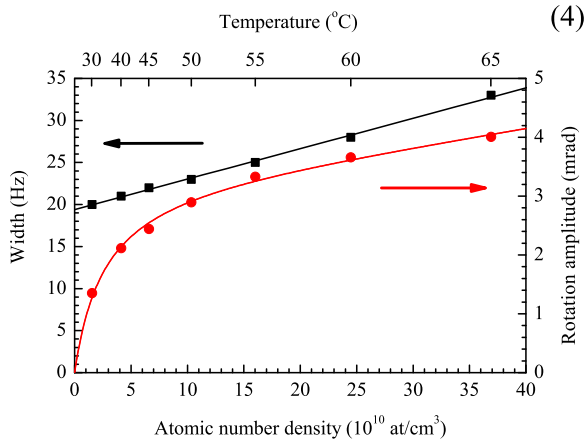}
   \caption{(Color online) Amplitude (red circles) and width (black squares)
   of the NMOR resonance vs. the atomic density and cell temperature. In
   the presented range, the width, and hence the relaxation rate
   of the ground-state coherences $\gamma$, increases linearly with
   the atomic density, while the amplitude dependence is fit to the phenomenological function
   $N/(1+N)e^{-N}$ (see text). The signals were measured with $I=3$~$\mu$W/mm$^2$.}
   \label{fig:TemperatureDependences}
  \end{figure}

Using Eq.~(\ref{eq:ConcentrationDependence}) we plot the density
dependences of the amplitude and width of the NMOR signal
(Fig.~\ref{fig:TemperatureDependences}).
The width dependence was fit with
\begin{equation}
\frac{\gamma_\text{c}}{2\pi}=a_\text{fit}N+\frac{\gamma_\text{in}}{2\pi},
\end{equation}
where $a_\text{fit}$ is the parameter proportional to the
collisional-relaxation cross-section and $\gamma_\text{in}$ denotes
the density independent relaxation rates. The fit yields
$a_\text{fit}=36.2(8)\times 10^{-12}$~Hz/cm$^{3}$ and
$\gamma_\text{in}=2\pi\times19.4(2)$~s$^{-1}$. Using
Eq.~(\ref{eq:TotalRelaxation}) with $R(I)\approx 0.2$
\cite{Okunevich1987}, one recalculates $a_\text{fit}$ into the
collisional-relaxation cross-section,
$\sigma_\text{exp}=2.79(6)\times10^{-14}$~cm$^2$. This value exceeds
by about 40\% the literature value \cite{Happer1972Review} of the
spin-exchange cross-section,
$\sigma_\text{se}=2.0(1)\times10^{-14}$~cm$^{-2}$. In fact, evidence
for excess surface relaxation depending on the atomic vapor density
has been reported also in other experiments with paraffin-coated
cells, see, for example,
Refs.~\cite{Budker2005Antirelaxation,Graf2005Relaxation}.

As shown in Fig.~\ref{fig:TemperatureDependences}, the amplitude of
the NMOR resonance depends on the atomic density in a nonlinear way.
For low densities the amplitude grows linearly with $N$, while for
higher densities the dependence levels off which is caused by the
rising light absorption in the cell, which is reflected by the $N
\exp(-N)/(1+N)$ dependence. In the context of NMOR, this effect was
previously studied in uncoated cells in
Ref.~\cite{Pustelny2005Density,Matsko2001Radiation}.

The sensitivity of the described magnetometric technique depends on
the slope of a central part of the NMOR resonance measured in the
modulation-frequency domain (see Sec.~\ref{sec:Noise}). To a good
approximation this slope is determined by a ratio of the signal
amplitude to its width. Thus for a given light intensity we take
this ratio as a magnetometer figure of merit.

\begin{figure}[h]
 \includegraphics[width=\columnwidth]{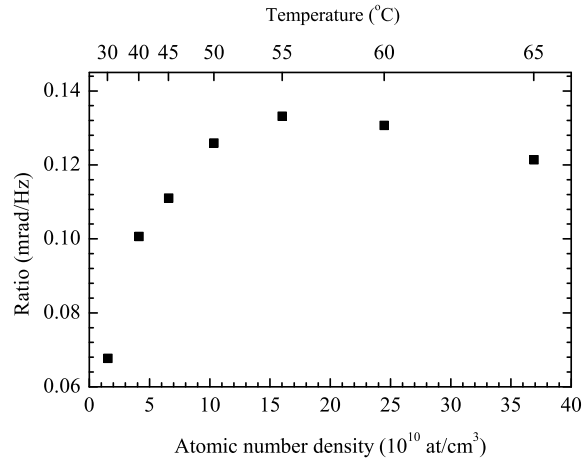}
 \caption{Amplitude-to-width ratio of the NMOR signal vs. the atomic number density and the temperature
 of vapor cell. For the fixed light intensity the ratio determines the sensitivity of the
 magnetic field measurements.}
 \label{fig:RatioTemperatureDependences}
\end{figure}

Using results presented in Fig.~\ref{fig:TemperatureDependences}, we
plot the amplitude-to-width ratio as a function of the atomic
density and cell temperature
(Fig.~\ref{fig:RatioTemperatureDependences}).
For low densities the ratio increases with concentration, but then
it saturates. For even higher concentrations the amplitude-to-width
ratio drops which results in reduced sensitivity of the
magnetic-field measurements. Existence of the maximum indicates the
optimal concentration for which the largest magnetometric
sensitivity of the method is achieved. The maximum arises due to
density-related relaxation and it occurs when the density-dependent
and density-independent relaxations become comparable (see
Fig.~\ref{fig:RatioTemperatureDependences}). Decrease of the
density-dependent relaxation rate would improve the method's
sensitivity. In our experimental conditions a distinct maximum
associated with the highest sensitivity was observed at $N\approx
1.6\times10^{11}$ atoms/cm$^3$ ($\sim$55$^\circ$C).

\section{Magnetic field tracking\label{sec:Tracking}}

In Sec.~\ref{sec:Results} we showed that the positions of the
high-field NMOR resonances in the magnetic field domain are
determined by the modulation frequency,
$B=\pm\hbar\Omega_\text{m}/2g\mu_\text{B}$. Thus, by controlling the
modulation frequency so that the resonance condition is fulfilled
one can track the varying magnetic field.

As discussed in Sec.~\ref{sec:Noise}, the sensitivity of the optical
magnetometer is determined by the slope of the resonance used for
field measurements. Therefore, from the point of view of the
magnetometer sensitivity, it is desirable to have strong and narrow
NMOR resonances. However, another important characteristic of the
magnetometer is its bandwidth, i.e., response time of the
magnetometer to a small change of a magnetic field. Therefore, the
choice of optimal conditions of the magnetometer requires
compromising between its sensitivity and bandwidth.

Analyzing the resonance signals, such as the one depicted in
Fig.~\ref{fig:ModulationFrequencySignal}, it can be shown that the
in-phase component is well suited for a magnetic field tracking. If
the resonance conditions are initially fulfilled, a change of the
magnetic field is reflected in a modification of the output signal,
negative if the field increases and positive if it decreases. Thus,
in the simplest realization of the magnetometer, the device can
operate with a feedback loop automatically adjusting the modulation
frequency to keep the signal level constant. However, a significant
disadvantage of this method is its limited dynamic range related to
the magnetic-field-dependent offset of the high-field resonances
(see inset in Fig.~\ref{fig:ModulationFrequencySignal}). Due to the
offset the magnetometer readout is burdened with a systematic error,
or for more significant field changes, for which the offset is
bigger than the amplitude of the signal, the magnetometer can
completely lose the field-tracking ability.

In order to avoid such errors one might take use of higher
derivatives of the signal or measure the NMOR signal at the higher
harmonics of the modulation frequency; instead we developed a
procedure that uses both components of the NMOR signal to track the
field. After a change of the field intensity the in-phase component
is first used to obtain information about a direction of the field
change and bring the system somewhere close to the resonance. It is
realized in the same manner as described above, i.e., the in-phase
readout is compared with a preset value and the modulation frequency
is coarsely modified to compensate the magnetic-field change. After
this adjustment, the magnetometer switches modes and starts to use
the quadrature component of the signal. At this stage the modulation
frequency is modified such that the minimum of the quadrature
component, and hence the field intensity is determined. After
finding a new position of the resonance, the in-phase reference
level is reset and this new value is used to track a successive
change of the field. Application of this procedure enabled a tenfold
increase of the dynamic range of the magnetic field measurement
(from 4~mG to 40~mG). At the current stage of the experiment, the
range is limited by technical issues.

\begin{figure}
   \begin{center}
    \includegraphics[width=\columnwidth]{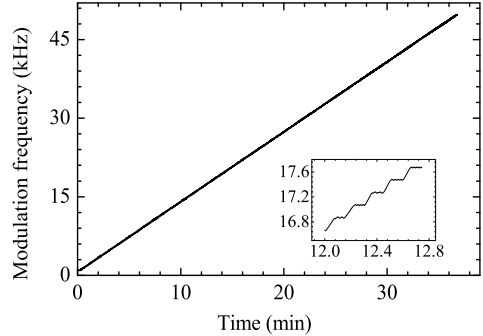}
   \end{center}
   \caption{Tracking signal of the magnetometer. Every 9~s the magnetic
   field was increased by $\sim$150~$\mu$G. After the change the magnetometer
   adjusted the modulation frequency to fulfill the resonance conditions
   (see inset).}
   \label{fig:Tracking}
  \end{figure}

In Fig.~\ref{fig:Tracking} the magnetometer tracking signal obtained
with the described method is shown in a range from 0.1~mG to ~40~mG.
For this measurement, every 9~s the field was increased by
$\sim$150~$\mu$G. After the field change, the magnetometer adjusted
itself to a new value using the algorithm described above. It is
worth noting that such arrangement enables measurement of static or
slowly-varying magnetic fields and remains immune to the oscillating
magnetic fields such as those associated with ac line.

\section{Performance and Sensitivity\label{sec:Noise}}

In this Section, we discuss the quantum limit on the sensitivity and
the obtained performance of the magnetometer.

\subsection{Quantum limit}

Sensitivity of optical magnetometers, such as the one described in
this paper, has an intrinsic quantum limit. In general, that limit
can be expressed as
\begin{equation}
    \delta B_\text{ql} = \sqrt{ \delta B_\text{at}^{2} + \delta B_\text{ph}^{2}},
    \label{eq:QuantumLimit}
\end{equation}
where $\delta B_\text{at}$ is the atomic shot-noise limit and
$\delta B_\text{ph}$ is the photon shot-noise limit.

The atomic shot-noise limit originates from fluctuations of the
number of atoms that contribute to the signal and depends on finite
lifetime of the light-induced atomic polarization. For alkali-vapor
magnetometers this limit is given by
\begin{equation}
    \delta B_\text{at} = \frac{\hbar}{g \mu_\text{B}} \times \sqrt{\frac{\gamma}{ V_\text{c} N}},
    \label{eq:AtomicShotNoise}
\end{equation}
where $V_\text{c}$ denotes the cell volume. Since the atomic
shot-noise limit is proportional to the square root of the
ground-state relaxation rate $\gamma$, a significant reduction of
that limit can be achieved in paraffin coated or buffer-gas vapor
cells. Moreover, analyzing Eq.~(\ref{eq:AtomicShotNoise}), one
obtains that higher sensitivity can be achieved with bigger cells
and in more dense media. However, as shown in
Sec.~\ref{sec:Results}, for higher atomic density relaxation due to
atomic collisions becomes an important factorf and $\gamma/N$ ratio
does not decrease with the concentration \footnote{Note that for
concentrations of atoms exceeding $10^{12}$~at/cm$^3$ ratio
$\gamma/N$ starts to drop with rising $N$. This effect is related
with limitation of the spin-exchange relaxation by frequent atomic
collisions. In fact, the up-to-date most sensitive magnetometer, the
so-called spin-exchange relaxation free (SERF) magnetometer,
exploits this mechanism for achieving sensitivity of a fraction of
10$^{-11}$ G/$\sqrt{\text{Hz}}$ in low-field measurements
\cite{Kominis2003Subfemtotesla}.}.

The second contribution to the quantum limit of the magnetometric
sensitivity is associated with the photon-shot noise in optical
polarimetry
\begin{equation}
    \delta \phi_\text{ph} = \frac{1}{2\sqrt{N_\text{ph}}},
    \label{eq:PhotonShotNoise}
\end{equation}
where $N_\text{ph}$ refers to the total number of photons incident
on the polarimeter. Expressing $N_\text{ph}$ as a function of the
light intensity, we derive formula for the photon-shot noise limit
on the magnetic field measurements
\begin{equation}
    \delta B_\text{ph} = \frac{\hbar}{g \mu_\text{B}} \times \frac{\gamma}{A_\text{N}} \times \frac{1}{2} \sqrt{\frac{2\pi\hbar c}{Ia\lambda \Delta{t}}},
    \label{eq:PhotonShotNoiseRecalculated}
\end{equation}
where $A_\text{N}$ is the NMOR signal amplitude, $I$ denotes the
average light intensity, $a=\pi (d/2)^2$ is the beam area with $d$
being the light-beam diameter, $\Delta t$ is the duration of the
measurement, and $c$ is the speed of light.

\begin{figure}[htb]
    \centering
        \includegraphics[width=\columnwidth]{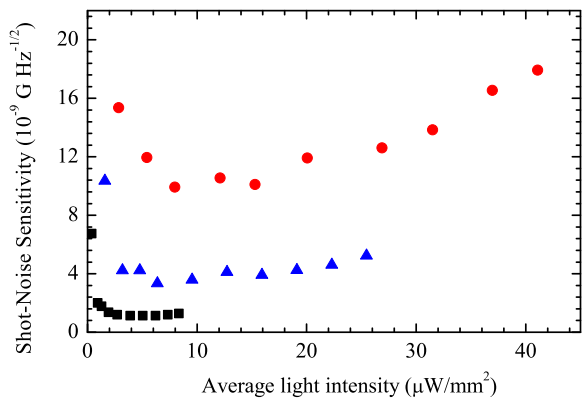}
    \caption{(Color online) Photon shot-noise limit on the magnetometric sensitivity $\delta B_\text{ph}$
    determined for three different modulation types: sine modulation (blue triangle),
    square wave modulations with 25\% (black squares) and 75\% (red circles) duty cycle,
    all with 100\% modulation depth and the same maximum light intensity.}
    \label{fig:ShotNoiseSensitivity}
\end{figure}

As discussed above, AM enables modification of the waveforms, duty
cycles, and modulation depths of light. A change of either of these
parameters is reflected in the amplitude $A_\text{N}$ and the width
$\gamma$ of the high-field resonance and, consequently, in the
sensitivity of the magnetic-field measurements. Using
Eq.~(\ref{eq:PhotonShotNoiseRecalculated}), the light-intensity
dependence of the photon shot-noise limit of the magnetic-field
sensitivity is plotted for three different modulation types used in
our experiment: the sine modulation and square-wave modulation with
25\% and 75\% duty cycles, all with 100\% modulation depths and the
same maximum light intensity (Fig.~\ref{fig:ShotNoiseSensitivity}).
The results, presented in Fig.~\ref{fig:ShotNoiseSensitivity}, show
that among these three modulation waveforms the highest photon
shot-noise limit on the magnetometric sensitivity can be achieved
for the square-wave modulation with 25\% duty cycle.

For the NMOR signals recorded with the described magnetometer we
have calculated the photon and atomic shot-noise limits on the
sensitivity being $\delta B_\text{ph}=3.9\times
10^{-10}$~G/$\sqrt{\text{Hz}}$ and $\delta B_\text{at}=3.2\times
10^{-11}$~G/$\sqrt{\text{Hz}}$, respectively, and the overall
quantum limit on the magnetometer sensitivity $\delta
B_\text{ql}\approx3.9\times 10^{-10}$~G/$\sqrt{\text{Hz}}$.

\subsection{Demonstrated sensitivity}

In order to find an intrinsic magnetometer sensitivity it is
necessary to determine the smallest magnetic-field change still
recognized by the device. In a perfectly shielded environment this
value is given by an amplitude of the recorded noise. However, in
real conditions, it is impossible to distinguish between the
fluctuations originating from the magnetic-field fluctuations
$\delta B_\text{f}$ from the ones related to different kinds of
noise $\delta B_\text{d}$ (shot-noise, polarimeter noise,
electronics noise, etc.). The latter are indeed limiting factors of
the magnetometer performance, while the former ones result from a
response of the magnetometer to magnetic-field changes and should
not be considered as performance limits. Therefore, for analysis of
the magnetometer sensitivity, it is practical to record a power
spectral density (PSD) of the rotation signals. PSD allows
determination of the signal-to-noise ratio ($S/N$) and provides
additional information about the system operation. For known $S/N$,
the sensitivity can be calculated as
\begin{equation}
    \delta B_\text{exp} = \frac{\hbar}{g \mu_\text{B}} \times \frac{\gamma_\text{m}}{S/N},
    \label{eq:NoiseMeasured}
\end{equation}
where $\gamma_\text{m}$ is the measured width of the NMOR resonance.

\begin{figure}[htb]
    \centering
        \includegraphics[width=\columnwidth]{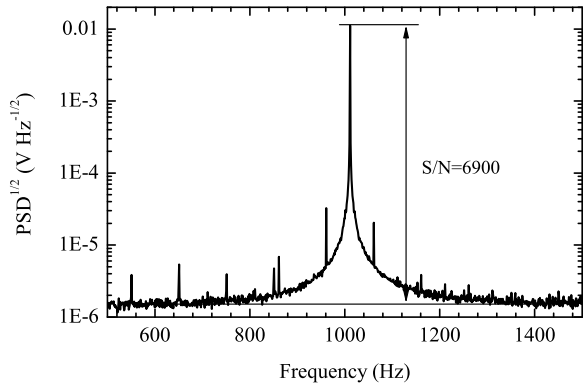}
    \caption{Square root of the power spectral density of the magnetometer signal.
    The central peak is the magnetometer response at twice the Larmor frequency
    at 0.72~mG. Sidebands are due to imperfect shielding of the
    $50$~Hz line. Broad pedestal of the signal is an artifact due to application
    of square windowing of the signal which precisely reproduces the signal
    amplitude but causes its broadening. The measurement was performed at $T\approx 55^\circ$C
    and $I=6.4$~$\mu$W/mm$^2$ with square-wave modulation of 41\% duty cycle.}
    \label{fig:Noise}
\end{figure}

Square root of the PSD of a typical signal provided by the operating
magnetometer is shown in Fig.~\ref{fig:Noise}. The central peak is
due to the Faraday rotation at twice the Larmor frequency, whereas
its nearest sidebands are related to magnetic field oscillating at
$50$~Hz, seen because of imperfect shielding of ac line. The value
of $S/N_\text{ext} \approx 6900$, depicted in Fig.~\ref{fig:Noise},
with the NMOR resonance width of $\approx 21$~Hz for the applied
pump beam intensity, corresponds to the measured magnetometer
sensitivity of $\delta B_\text{exp} = 4.3\times
10^{-9}$~G/$\sqrt{\text{Hz}}$.

The demonstrated sensitivity of the magnetometer is about an order
of magnitude lower than its shot-noise limit. This is related to the
fact that apart from the quantum limit, there are other factors
which more severely limit the magnetometer sensitivity. One such
factor is the atomic density. According to
Ref.~\cite{Auzinsh2004Can}, the strongest NMOR signals are observed
in a medium with an optical depth close to unity, i.e., when each
photon is scattered by atoms once. In room-temperature rubidium
vapors contained in cells of small dimensions the optical depth is
lower than unity, so it is necessary to heat the cell to increase
the density. However, along with the rise of the atomic density, the
adverse density-dependent relaxation becomes faster and broadens the
NMOR resonances. It is thus necessary to compromise between the
width and amplitude of the rotation signal when working with small
cells. In our experiment in which the 2-cm long vapor cell was used,
the best compromise was achieved for optical depth about 0.35 that
is about 3 times smaller than in the case of the large cell setup
with optimized shot-noise sensitivity. Additional suppression of the
sensitivity is caused by the light scattering losses and electronic
component noise.

\section{Conclusions\label{sec:Conclusions}}

We have demonstrated the all-optical magnetometric technique which
allows measurements of slowly varying magnetic fields in the range
of 40~mG with the sensitivity of
4.3$\times$10$^{-9}$~G/$\sqrt{\text{Hz}}$. We discussed the quantum
limit of the method associated with the atomic and photon
shot-noise. The shot-noise analysis showed  that our experimental
technique still leaves room for improvement.

Work towards extension of the dynamic range of magnetic field
measurements to higher field (ultimately the Earth magnetic field)
and increasing of the sensitivity of the method is in progress. In
particular, we concentrate on development of the self-oscillating
magnetometers \cite{Higbie2006AllOptical,Pustelny2006AllOptical}. In
that technique the NMOR signal is used for modulation of pumping
light. This realization of the experiment enables measurements of
magnetic fields varying with a frequency up to 1 kHz and offers
substantial simplification of the setup which is important from a
point of view of commercialization of the technique.

\begin{acknowledgements}
The authors would like to express their gratitude to Dmitry Budker,
Andrzej Ku\l ak, Stanis\l aw Micek, and Simon Rochester for
stimulating discussions and J\'ozef Flaga and Stanis\l aw Pajka for
their technical assistance in realization of the experiment. This
work has been supported by Polish Ministry Of Science and Higher
Education grant \# N N505 0920 33. One of the authors (S.P.) is a
scholar of the Foundation for Polish Science.
\end{acknowledgements}

\end{document}